\documentclass[a4paper,11pt]{article}

\usepackage{graphicx}
\usepackage{amsmath}
\usepackage{amsfonts}
\usepackage{amssymb}
\usepackage{units}
\usepackage{epsf}
\usepackage{bbm}   
\newcommand{\be}{\begin{equation}}
\newcommand{\ee}{\end{equation}}
\newcommand{\ba}{\begin{align}}
\newcommand{\ea}{\end{align}}

\newcommand{\braket}[2]{\ensuremath{\langle \, #1 \, |\, #2 \, \rangle}}

\newcommand{\ket}[1]{\ensuremath{| \, #1 \, \rangle }}

\newcommand{\1}{\ensuremath{\mathbbm{1}}}

\begin{document}
\title{An upper limit on CP violation in the $B^0_s-\bar{B}^0_s$ system
\author{Ch. Berger \thanks{e-mail: berger@rwth-aachen.de}
\\{\small I. Physikalisches Institut, RWTH Aachen University, Germany} \and
L. M. Sehgal\thanks{e-mail: sehgal@physik.rwth-aachen.de}
\\{\small Institute for Theoretical Particle Physics and Cosmology,}
\\{\small RWTH Aachen University, Germany}}
}

\date{}

\maketitle

\begin{abstract}
In a previous publication we noted that the time dependence of an
incoherent $B^0-\bar{B}^0$ mixture undergoes a qualitative
change when the magnitude of CP violation $\delta$ exceeds a critical value. 
Requiring, on physical grounds, that the system evolve from an initial incoherent
state to a final pure state in a monotonic way, yields a new upper limit for
$\delta$. The recent measurement of the wrong charge semileptonic asymmetry
of $B_s^0$ mesons presented by the D0 collaboration
is outside this bound by one standard deviation. If this result is confirmed
it implies the existence of a new quantum mechanical oscillation phenomenon.
\end{abstract}

In a previous paper~\cite{CPVpaper} we studied the time evolution
of an incoherent $B^0-\bar{B}^0$ mixture as a function of the strength
of the CP-violating parameter 
\be
\delta=\braket{B_L}{B_S}=\frac{|p|^2-|q|^2}{|p|^2+|q|^2}\enspace ,
\label{asl1}\ee
where $B_L$ and $B_S$ denote the long-lived and short-lived
eigenstates of the system which can be expressed as  superpositions
of $B^0$ and $\bar{B}^0$ with the coefficients $p,q$:
\begin{eqnarray}
 \ket{B_L}&=&\frac{1}{\sqrt{|p|^2+|q|^2}}(p\ket{B^0}-q\ket{\bar{B}^0})\nonumber\\
\ket{B_S}&=&\frac{1}{\sqrt{|p|^2+|q|^2}}(p\ket{B^0}+q\ket{\bar{B}^0})
\label{asl1a}
\end{eqnarray}

The density matrix, in the $B^0-\bar{B}^0$ basis, was written as 
\be
 \rho(t) = \frac{1}{2} N(t) \left[\1 + \vec{\zeta}(t) \cdot \vec{\sigma} \right]
  \enspace ,\label{asl2}
 \ee
where the normalisation function $N(t)$ and the Stokes vector $\vec{\zeta}(t)$
had initial values $N(0)=1$, $\vec{\zeta}(0)=0$ respectively.
By explicit calculation one finds
\be N(t) = \frac{1}{2(1-\delta^2)} \left[ e^{-\gamma_S t} + e^{-
\gamma_L t} - 2 \delta^2 e^{ - \frac{1}{2} ( \gamma_S +
\gamma_L)t} \cos\Delta mt \right]\enspace .\label{asl3}
\ee
Furthermore, the magnitude of the Stokes vector is found to have a 
remarkable relation to $N(t)$: 
\be
|\vec{\zeta}(t)|=\left[1- \frac{1}{N(t)^2} e^{- ( \gamma_S + \gamma_L)t}\right]^{\frac{1}{2}}
\enspace .\label{asl4}\ee
It should be stressed that Eqs.~(\ref{asl3},\ref{asl4})
are valid for any system such as $ K^0-\bar{K}^0$, $B_d^0-\bar{B}_d^0$
or $B_s^0-\bar{B}_s^0$ satisfying the Lee-Oehme-Yang equation of 
motion~\cite{LOY}. In particular the relation~(\ref{asl4})
between $|\vec{\zeta}(t)|$ and $N(t)$
does not involve the parameters $\delta$, $\Delta\gamma$
or $\Delta m$ where $\Delta\gamma$ and $\Delta m$ denote the 
differences in the decay
widths and masses of the eigenstates. Our results are independent of the sign
of $\Delta\gamma/\Delta m$.

An upper bound on $\delta^2$ arises from the requirement that the normalisation function
$N(t)$ should be a monotonically decreasing function of time,
i.e. $dN/dt<0$ for all $t$. This leads to the constraint
\be
\delta^2 \leq  \left( \frac{\gamma_S \gamma_L}{
  \left( \gamma_S+ \gamma_L\right)^2\!\!/4 + \Delta m^2}\right)^{\frac{1}{2}}
\label{asl5}\enspace .\ee
In fact, a stronger constraint is obtained by demanding that the 
time-dependent norm of an arbitrary pure state
$\alpha\ket{B^0}+\beta\ket{\bar{B}^0}$ decrease monotonically with
time. This bound was obtained by Lee and Wolfenstein~\cite{LW} and
by Bell and Steinberger~\cite{BSt}, and is referred to as the
unitarity bound
\be
\delta^2_{\rm unit} \leq  \left( \frac{\gamma_S \gamma_L}{
  \left( \gamma_S+ \gamma_L\right)^2\!\!/4 + \Delta m^2}\right)
\label{asl6}\enspace .\ee

In~\cite{CPVpaper} we derived a complementary bound from the requirement
that the function $|\vec{\zeta}(t)|$ evolve from its
initial value $|\vec{\zeta}(0)|=0$ to its final value
$|\vec{\zeta}(\infty)|=1$ (representing the long-lived pure state
$B_L$) in a monotonic way. This amounts to the assumption that the
transformation of the $B^0-\bar{B}^0$ mixture from its incoherent
(high entropy) beginning to its final pure (low entropy)
end-state is unidirectional in time. 
The common requirement of monotonicity for $N(t)$ and
$|\vec{\zeta}(t)|$ is equivalent to the statement that both of these
observables should behave as arrows of time~\cite{CPVpaper}.

The monotonicity condition $d|\vec{\zeta}(t)|/dt\geq 0$ can be
expressed as a condition on $N(t)$ making use of the
relation~(\ref{asl4}). As shown in~\cite{CPVpaper}
this implies an upper bound
\be
\delta^2 \leq \frac{1}{2} \left( \frac{ \Delta \gamma }{\Delta m}\right)
  \sinh \left( \frac{3\pi}{4} \; \frac{ \Delta \gamma }{\Delta m}\right)
\enspace. \label{asl7}
\ee
which reduces for small $\Delta \gamma /\Delta m$, as found in the
$B^0-\bar{B}^0$ system, to
\be
|\delta|\leq\sqrt{\frac{3\pi}{8}}\left|\frac{\Delta\gamma}{\Delta m}\right|
\enspace . \label{asl8}
\ee

A useful experimental observable is the  ``wrong charge'' semileptonic asymmetry
defined (with $q=d,s$) as
\be
a^q_{sl}=\frac{\Gamma(\bar{B}^0_q\to \mu^+X)-\Gamma(B^0_q\to \mu^-X)}{\Gamma(\bar{B}^0_q\to \mu^+X)+\Gamma(B^0_q\to \mu^-X)}
\label{asl9a}  
\ee
which is directly related to the  CP violating parameter (for recent reviews see~\cite{aslref}) 
\be
a^q_{sl}=\frac{1-\left|\frac{q}{p}\right|^4}{1+\left|\frac{q}{p}\right|^4}
=\frac{2\delta}{1+\delta^2}
\label{asl9}
\ee
reducing to $a^q_{sl}=2\delta$ for small $\delta$.

\begin{figure}[h]
\begin{center}
\includegraphics[width=0.74\textwidth]{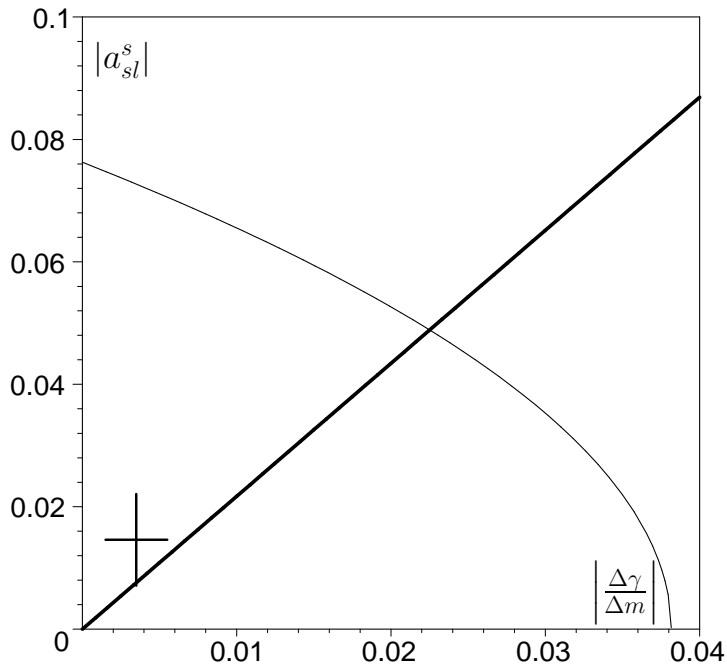}
\end{center}

\caption{{\small Constraints on $|a^s_{sl}|$
 in the $|a^s_{sl}|-|\Delta \gamma/\Delta m|$ plane
resulting from unitarity and monotonicity
of $|\zeta(t)|$ for the $B_s^0-\bar{B}_s^0$ system. The
thin line represents the unitarity bound (\ref{asl6}) with $\Delta m/\gamma_S=26.2$ and the thick
line our new bound
evaluated from (\ref{asl7}). The cross represents the
 experimental result of the D0 experiment for $|a_{sl}^s|$ with the horizontal error bar
indicating the uncertainty of $|\Delta\gamma/ \Delta m|$}}
 
\label{aslfig1}\end{figure}

The ``coherence bound'' for $|a^s_{sl}|$ calculated from (\ref{asl7})   is plotted
in Fig.~\ref{aslfig1} versus $|\Delta \gamma /\Delta m|$ together with the unitarity bound obtained 
from
(\ref{asl6}). It is immediately seen that for $|\Delta \gamma /\Delta m|\to 0$
the coherence bound is the more effective of the two, providing stringent upper limits.

In a recent analysis of like-sign dimuon pairs in the D0 experiment~\cite{D0} an asymmetry 
was measured between $\mu^+\mu^+$ and $\mu^-\mu^-$ final states,
\be
A_{sl}^b=\frac{N_b^{++}-N_b^{--}}{N_b^{++}+N_b^{--}}=-(0.957\pm 0.251\pm 0.146)\%
\enspace .\label{asl10}
\ee  
This asymmetry was determined to be a linear combination of the semileptonic
wrong charge asymmetry associated with $B_d$ and $B_s$ mesons
\begin{equation}
 A_{sl}^b=(0.506\pm 0.043)a_{sl}^d+(0.494\pm 0.043)a_{sl}^s
\enspace . \label{asl11a}
\end{equation}
 
Data from $B_d$-factories~\cite{Barb} provide an estimate of the asymmetry parameter
$a_{sl}^d$
\begin{equation}
a_{sl}^d=-(0.47\pm 0.46)\%
\enspace . \label{asl11}
\end{equation}
Combining this information with Eq.~(\ref{asl11a}) the D0 experiment
has extracted a value for the parameter $a_{sl}^s$ which is negative in sign and
has the modulus
\begin{equation}
 |a_{sl}^s|=(1.46\pm 0.75)\%\enspace .
\label{asl12}
\end{equation}
This result is included in Fig.~\ref{aslfig1}. The horizontal error bar reflects
the uncertainty in the measured value~\cite{PDG}
 of $|\Delta\gamma/ \Delta m|=0.0035\pm 0.002$.  
The figure shows that the experimental value of $|a_{sl}^s|$ is clearly
outside the allowed region violating the 
coherence bound by about one standard deviation.

If the Hamiltonian governing the time-dependence of the $B^0-\bar{B}^0$ system
is written as a $2\times2$ matrix $M_q-\imath\Gamma_q/2$,
where $M_q$ and $\Gamma_q$ are hermitian, the origin of CP violation resides in the phase
\begin{equation}
\Phi_q=\arg\left(-\frac{M_q^{12}}{\Gamma_q^{12}}\right)
\enspace .\label{asl14}\end{equation}
In terms of this phase the wrong charge asymmetry parameter is given by
\begin{equation}
a_{sl}^q=\frac{\Delta \gamma}{\Delta m}\tan \Phi_q
\label{asl13}\end{equation}
neglecting higher orders in $\Delta\gamma/\Delta m$.
The linear dependence of (\ref{asl13}) on $\Delta\gamma/ \Delta m$ allows an immediate combination
with (\ref{asl8}) leading to the constraint
\begin{equation}
 \left|\tan\Phi_s\right|\leq \sqrt{\frac{3\pi}{2}}=2.17\enspace .
\label{asl15}\end{equation}
Note that the standard model value for $\Phi_s$ is very small~\cite{Beneke} $\Phi_s^{SM}=0.24^\circ$.
The bound~(\ref{asl15}) may serve as a new constraint on theories that
go beyond the standard model (see for example, the papers
in~\cite{BSMpapers}). It should be stressed that variations of $\Phi_s$ must in
any case respect the unitarity limit shown in  Fig.~\ref{aslfig1}.
The upper limit $a_{sl}^s <7.6\%$ implies the unitarity bound 
$|\tan\Phi_s|< 22$.

Our analysis based on elementary quantum mechanical reasoning 
(requiring no assumption about the Hamiltonian
other than CPT invariance) shows
interesting new bounds on the CP violating parameters of the
$B^0-\bar{B}^0$ system. It is, however, not granted
that nature respects the monotonicity bound following from~(\ref{asl7}).
In this case the D0 result -- if true -- implies the existence of 
a new type of quantum mechanical oscillation.

\begin{figure}[h]
\begin{center}
\includegraphics[width=0.74\textwidth]{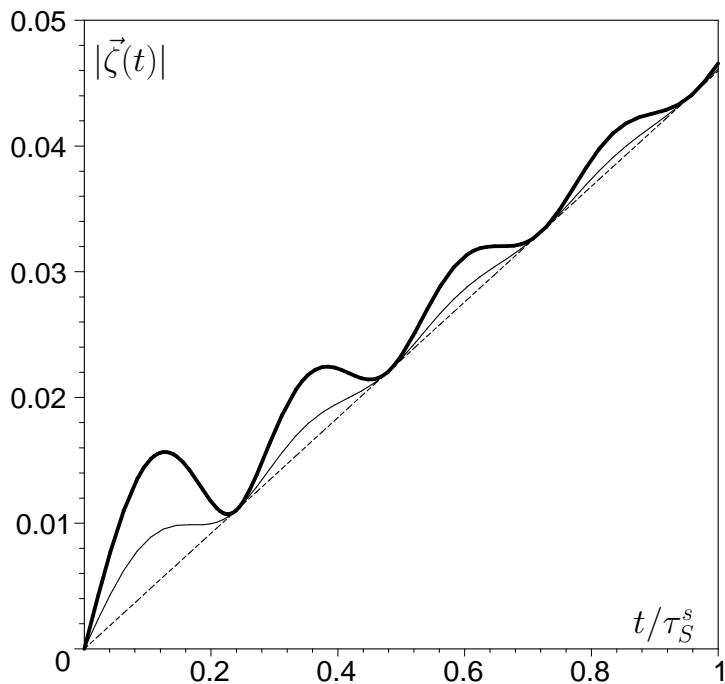}
\end{center}

\caption{{\small  Plot of $|\vec{\zeta}(t)|$ versus $t$ in units of
the lifetime $\tau_S^s$. The thick line is calculated for
the nominal D0 value of $\delta=0.0073$. The strictly monotonic
dashed line is obtained for the standard model value of $\delta$.
The thin line represents the behaviour
of $|\vec{\zeta}(t)|$ at the critical value $\delta_{\rm crit}=0.0038113$ i.e.
the transition between monotonic and nonmonotonic regimes.}}
 
\label{aslfig2}\end{figure}
To exhibit this new phenomenon we examine in 
Fig.~\ref{aslfig2} the characteristics of $|\vec{\zeta}(t)|$ as a function of time
in units of $\tau_S^s$, the lifetime of the short lived $B^0_s$ meson.
The thick line calculated for the nominal D0 value of $\delta=0.0073$ exhibits two
clear oscillations within the first half of the  $B^0_s$ lifetime. By contrast  the
dashed line evaluated for the standard model with $\delta \approx 10^{-5}$
is strictly monotonic and in the chosen range of $t/\tau^s_S$ even linearly increasing in time.
At the critical value $\delta_{\rm crit}=0.0038113$ (found by numerical
evaluation of the condition $d|\zeta(t)|/dt\geq 0$) 
the evolution of $|\vec{\zeta}(t)|$
in Fig.~\ref{aslfig2} follows the middle curve which marks
the line of transition between the monotonic and nonmonotonic regimes. 

If the D0 result is confirmed it should be possible, in principle, to insert precise data on $N(t)$
into~(\ref{asl4}) in order to reveal the coherence oscillations shown in the upper curve
of Fig.~\ref{aslfig2}. Since the whole effect is induced by the term proportional to $\delta^2$
in~(\ref{asl3}), the precision required would be at least $\cal{O}$ $\!\!(10^{-4})$, which may be 
beyond reach. It is interesting, nevertheless, that precise knowledge of $N(t)$ would allow one
to probe the coherence function $\zeta (t)$, and determine whether or not the evolution of 
coherence in a $B_s^0-\bar{B}_s^0$ system follows an arrow of time. 
The D0 asymmetry, taken at face value, 
suggests that this arrow is broken.

Note added in proof: Since submission of this paper the Heavy Flavor Averaging Group
has reanalyzed~\cite{HFAG} the available published~\cite{D0,D02,CDF1}
and unpublished~\cite{CDF2} experimental data
related to $a_{sl}^s$ and
extracted an average value of $a_{sl}^s= (-0.85\pm 0.58)\%$. The central
value is much closer to our limit. In view of the large error a precise determination
of  $a_{sl}^s$ is eagerly awaited. 


\end{document}